\documentclass[%
preprint, 
 amsmath,amssymb,
 aps, physrev,
]{revtex4-2}

\usepackage{graphicx}
\usepackage{dcolumn}
\usepackage{bm}
\usepackage{xcolor}%
\usepackage{siunitx}
\usepackage{orcidlink}

\begin{document}


\title{Self-consistent treatment of Intra Beam Scattering, betatron coupling, and vertical dispersion in fourth generation light sources}

\author{Sébastien Joly\orcidlink{0009-0003-3174-9735}}
 \email{sebastien.joly@helmholtz-berlin.de}
\affiliation{Helmholtz-Zentrum Berlin für Materialien und Energie GmbH (HZB), Berlin, Germany}

\author{Jonas Kallestrup\orcidlink{0000-0002-0721-9393}}
\affiliation{Paul Scherrer Institut, CH-5232 Villigen PSI, Switzerland}

\author{Félix Soubelet\orcidlink{0000-0001-8012-1440}}
\affiliation{CERN, Geneva, Switzerland}


\date{\today}

\begin{abstract}

The X-ray brightness delivered by fourth-generation light sources strongly depends on the electron beam current and transverse emittance. Reaching higher brilliance and lower emittances are increasingly limited by intra beam scattering, particularly at low and medium beam energies, where low emittances combined with high beam currents result in large phase-space densities.
Increasing the vertical emittance through betatron coupling is commonly employed to mitigate intra beam scattering by relaxing the phase-space density. However, the redistribution of damping partition numbers due to coupling, the presence of vertical dispersion, and consequently their impact on the balance between synchrotron radiation and intra beam scattering are often neglected.

In this work, we develop a self-consistent Ordinary Differential Equations-based framework to describe both the steady-state and time evolution of three-dimensional beam emittances in the simultaneous presence of synchrotron radiation, quantum excitation, betatron coupling, vertical dispersion, and intra beam scattering; allowing for realistic damping partition numbers. The model consistently accounts for the modification of synchrotron radiation damping rates and intra beam scattering growth rates arising from betatron coupling.

Application to the BESSY III lattice demonstrates that damping partition redistribution and optics modifications significantly influence the equilibrium emittances. A systematic comparison of vertical emittance generation via a transverse feedback-generated excitation, betatron coupling, and vertical dispersion highlights the trade-offs between horizontal emittance reduction and operational constraints. 

\end{abstract}

\maketitle










\section{Introduction}

Synchrotron light sources are electron storage rings that exploit the intense, high-brightness X-ray radiation emitted by relativistic electrons in magnetic fields. These facilities support a broad range of high-impact applications in materials science, chemistry, energy research, life sciences, and related fields.
The brightness of the emitted radiation depends on the type of magnetic elements that bend the electron beam trajectory and on the electron beam properties. Fourth-generation light sources aim to enhance brightness primarily by reducing the transverse emittance to the sub-nanometer regime, where the electron beam size approaches the diffraction limit. 
One effective strategy to achieve ultra-low emittances in fourth-generation light sources is the usage of a Multi-Bend Achromat (MBA) lattice incorporating combined-function magnets~\cite{leemann_beam_2009, raimondi_extremely_2023, liu:ipac13-tupwo001}. Combined-function magnets, such as reverse bends, modify the damping partition numbers by increasing the horizontal partition $J_x$ and decreasing the longitudinal partition $J_z$. A larger $J_x$ leads to stronger horizontal radiation damping and consequently to a lower equilibrium horizontal emittance.
This design was adopted for BESSY~III~\cite{goslawski:ipac24-tupg28, goslawski_non-standard_2025}, a \SI{2.5}{\GeV} fourth-generation light source meant to operate with a transverse natural emittance of approximately~\SI{100}{\pm \radian} and a nominal beam current of \SI{300}{\mA}.

However, an intense beam current together with a low emittance results in a dense phase space volume, enhancing the impact of Coulomb scattering. Small-angle Coulomb scatterings give rise to the Intra Beam Scattering (IBS) effect, leading to a redistribution of the bunch phase space and subsequently an increase of the beam emittance~\cite{smaluk_electron_2024, agapov_beam_2026} at the operated energies. The IBS strength increases sharply as the beam emittance and energy decrease. Thus, it is currently the most limiting factor for reaching smaller emittances in low and medium-energy (up to a few GeV) synchrotron light sources.

The two main strategies to mitigate the impact of IBS are, first, to lengthen the bunch using a Higher Harmonic Cavity (HHC)~\cite{leemann_interplay_2014, gubaidulin_interaction_2025, bassi:ipac23-wepl141}, thereby reducing the peak longitudinal charge density, and second, to increase the vertical emittance via betatron coupling~\cite{cortes:ipac25-mops045, joly:ipac25-wepm024}, which lowers the vertical phase-space density. However, the redistribution of the damping partition numbers induced by betatron coupling, and its consequences on the steady-state emittances in the presence of IBS is frequently neglected in the literature. 
The change in the damping partitions modifies the synchrotron radiation (SR) damping rates and therefore impacts the equilibrium between SR and IBS. Besides, the characteristic timescales associated with SR, IBS, and betatron coupling are comparable. A self-consistent treatment of this interplay is therefore essential for an accurate prediction of steady-state emittances.

In this paper, we investigate the mitigation of IBS through the introduction of vertical emittance, with a particular focus on fourth-generation light sources. We first introduce an Ordinary Differential Equations (ODE) based framework to compute equilibrium emittances in the presence of quantum excitation (QE), SR, betatron coupling, and vertical dispersion for arbitrary damping partition numbers. The resulting solutions are compared with analytical expressions available in the literature and benchmarked against simulations performed with pyAT~\cite{pyat_2017}.
Then, the approach is extended to include IBS, enabling the calculation of the full time evolution of the three-dimensional emittances. The predicted steady-state emittances are then compared with the current Xsuite~\cite{xsuite} implementation, and the observed differences are discussed.
Finally, the ODE-based framework is applied to the BESSY III lattice to evaluate the steady-state emittances obtained for different sources of vertical emittance, namely betatron coupling, vertical excitation, and vertical dispersion.

\section{Tools to calculate the steady-state emittances with IBS}

The IBS was first described analytically by Piwinski~\cite{Piwinski:400720} and independently by Bjorken and Mtingwa~\cite{Bjorken:140304}, who characterized its effect in terms of emittance growth rates. Subsequent developments either significantly accelerated the computation of these growth rates~\cite{nagaitsev_intrabeam_2005} by neglecting the vertical dispersion or simplified their expressions using the high energy approximation~\cite{bane_simplified_2002, kubo_intrabeam_2005}.
Based on these IBS growth rates, the steady-state emittances in the presence of various effects can be calculated using the ODEs approach, as implemented in codes such as ZAP~\cite{zisman_zap_1986}, elegant~\cite{borland_elegant_2000}, and Xsuite. The validity of this approach is limited to the emittance growth of the bunch's core, under the assumption that the particle distribution remains Gaussian.
The growth rates can also be used to generate semi-analytical IBS kicks, as implemented in elegant~\cite{borland_elegant_2000}, Xsuite, and mbtrack2~\cite{gamelin:ipac21-mopab070}. The kicks are derived from the IBS growth rates assuming a Gaussian distribution, which reasonably approximates most observed bunch distributions.
Two additional approaches can be mentioned: the Monte Carlo method used in SIRE~\cite{vivoli:ipac10-wepe090} and CMAD-IBStrack~\cite{demma:ipac11-wepc105,pivi:pac07-thpas066,pivi:ipac12-weppr091}, and the beam envelope matrix formalism implemented in SAD~\cite{kubo_intrabeam_2001}.

In the following, the ODEs approach is extended to account self-consistently for betatron coupling. Before doing so, we introduce the definitions of projected and mode emittances and discuss their differences in the presence of betatron coupling.

\section{Projected and mode emittances in the presence of betatron coupling}

The transverse bunch distribution is described by the $4\times4$ covariance matrix $\Sigma$, which describes the four-dimensional phase space through its second-order moments and whose off-diagonal blocks represent the coupling terms. It results from linear transport, radiation damping, and quantum excitation over one turn:

\begin{equation}
\Sigma =
\begin{bmatrix}
\langle x^2 \rangle & \langle x x' \rangle & \langle x y \rangle & \langle x y' \rangle \\
\langle x x' \rangle & \langle x'^2 \rangle & \langle x' y \rangle & \langle x' y' \rangle \\
\langle x y \rangle & \langle x' y \rangle & \langle y^2 \rangle & \langle y y' \rangle \\
\langle x y' \rangle & \langle x' y' \rangle & \langle y y' \rangle & \langle y'^2 \rangle
\end{bmatrix}.
\end{equation}

Based on the $\Sigma$ matrix, two complementary emittance definitions can be introduced.
The eigen (or mode) emittances $\mathcal{E}_{u,v}$ are the invariants of $\Sigma$ under symplectic transport and therefore do not depend on the longitudinal position $s$ along the ring. In contrast, the projected emittances $\varepsilon_{x,y}$ are obtained from the determinants of the horizontal and vertical $2\times2$ submatrices of $\Sigma$, neglecting the cross-correlation terms. As a result, projected emittances generally vary along the ring in the presence of coupling. The two definitions become equivalent only in the absence of coupling. In the remainder of this paper, the term emittance refers to the projected emittances unless stated otherwise.

Using the second-order moment of the particle distributions' mapping approach developed in~\cite{kuske, kuskepeter_bloch}, the projected emittances in the presence of betatron coupling (neglecting the sum resonance), QE, and SR can be expressed as~\cite{kuske2}:

\begin{equation}
\begin{split}
    \varepsilon_x & = \varepsilon_0 \frac{|C^-|^2 \left(1 + \frac{\alpha^{SR}_x}{\alpha^{SR}_y} \right) + 4\Delta^2 + \left[ \frac{T_0}{\pi} \left(\alpha^{SR}_x + \alpha^{SR}_y\right)\right]^2}{|C^-|^2 \left( \frac{\alpha^{SR}_y}{\alpha^{SR}_x} + \frac{\alpha^{SR}_x}{\alpha^{SR}_y} + 2 \right) + 4\Delta^2 + \left[ \frac{T_0}{\pi} \left(\alpha^{SR}_x+ \alpha^{SR}_y\right)\right]^2}, \\
    \varepsilon_y & = \varepsilon_0 \frac{|C^-|^2 \left(1 + \frac{\alpha^{SR}_x}{\alpha^{SR}_y} \right)}{|C^-|^2 \left( \frac{\alpha^{SR}_y}{\alpha^{SR}_x} + \frac{\alpha^{SR}_x}{\alpha^{SR}_y} + 2 \right) + 4\Delta^2 + \left[ \frac{T_0}{\pi} \left(\alpha^{SR}_x + \alpha^{SR}_y\right)\right]^2},
\end{split}
\label{eq:proj_eps_peter}
\end{equation}

where $\varepsilon_0$ is the natural emittance (equal to $\varepsilon_{x,0}$ when the vertical dispersion is negligible), $\alpha_{x,y}^{SR} = 1 / \tau_{x,y}^{SR}$ are the horizontal and vertical synchrotron radiation damping rates, $T_0$ is the revolution period, $\Delta$ is the fractional tune separation, and $|C^-|$ is the coupling coefficient~\cite{guignard_betatron_1995}, defined as:

\begin{equation}
    \left| C^- \right| = \frac{1}{2 \pi} \left| \oint ds k_s(s) \sqrt{\beta_x(s) \beta_y(s)} e^{i \left(\phi_x(s) - \phi_y(s) - 2\pi \Delta \frac{s}{L} \right)} \right|.
\label{eq:C-}
\end{equation}

Here, $k_s$ denotes the skew quadrupole gradient, $\beta_{x,y}$ are the horizontal and vertical beta functions, $\phi_{x,y}$ are respectively the horizontal and vertical phase advances, and $L$ is the ring circumference.
The coupling coefficient approaches the on-resonance coupling coefficient, $|C^-_0|$, in the limit $\Delta \rightarrow 0$.

\section{Time evolution of the emittance with betatron coupling}

While Eq.~\eqref{eq:proj_eps_peter} provides the equilibrium transverse emittances, their time evolution can be described using a system of coupled differential equations~\cite{Lee}, simultaneously including radiation damping and transverse emittance exchange:

\begin{equation}
\begin{split}
    \frac{d \varepsilon_x}{dt} = & -\alpha^C \left(\varepsilon_x - \varepsilon_y \right) - 2\alpha^{SR}_x \left( \varepsilon_x - \varepsilon_{x,0}\right), \\
    \frac{d \varepsilon_y}{dt} = & -\alpha^C \left(\varepsilon_y - \varepsilon_x \right) - 2\alpha^{SR}_y \left( \varepsilon_y - \varepsilon_{y,0}\right).
\end{split}
\label{eq:heuristic_eps}
\end{equation}

Only the difference resonance contributes to the emittance exchange in this model, represented by the term proportional to $\alpha^C$, the coupling rate. A full treatment including both the sum and difference resonances is given in Appendix~\ref{app:A}. 
Accounting solely for the difference resonance is well justified, since in most storage rings the sum resonance is significantly smaller than the difference resonance due to the chosen working point, thus its contribution to the emittance is negligible.

By solving for the steady-state solutions of the coupled system of differential equations, imposing $\frac{d \varepsilon_x}{dt} = \frac{d \varepsilon_y}{dt} = 0$, the equilibrium emittances are obtained as:

\begin{equation}
\begin{split}
\varepsilon_x & = \frac{\varepsilon_{x,0} \alpha^C \alpha_x^{SR} + 2 \varepsilon_{x,0} \alpha_x^{SR} \alpha_y^{SR} + \varepsilon_{y,0} \alpha^C \alpha_y^{SR}}{\alpha^C \alpha_x^{SR} + \alpha^C \alpha_y^{SR} + 2 \alpha_x^{SR} \alpha_y^{SR}}, \\
\varepsilon_y & = \frac{\varepsilon_{x,0} \alpha^C \alpha_x^{SR} + \varepsilon_{y,0} \alpha^C \alpha_y^{SR} + 2 \varepsilon_{y,0} \alpha_x^{SR} \alpha_y^{SR}}{\alpha^C \alpha_x^{SR} + \alpha^C \alpha_y^{SR} + 2 \alpha_x^{SR} \alpha_y^{SR}}.
\end{split}
\label{eq:eps_ss}
\end{equation}

The above formulas are valid for a beam with a finite vertical emittance arising from vertical dispersion through the term $\varepsilon_{y,0}$.
It is convenient to define the emittance ratio $\kappa = \varepsilon_y / \varepsilon_x$. Then, using Eq.~\eqref{eq:eps_ss}, $\alpha^C$ can then be isolated as:

\begin{equation}
    \alpha^C = \frac{2 \alpha_x^{SR} \alpha_y^{SR} \left( \varepsilon_{y,0} - \kappa \varepsilon_{x,0} \right)}{(\kappa - 1) \left( \varepsilon_{x,0} \alpha_x^{SR} + \varepsilon_{y,0} \alpha_y^{SR} \right)}.
\end{equation}

A further simplification is obtained by neglecting the vertical emittance, $\varepsilon_{y,0} \approx 0$, as is typically the case in the absence of vertical dispersion. Defining $\varepsilon_0 = \varepsilon_{x,0}$ and substituting the steady-state emittances from Eq.~\eqref{eq:proj_eps_peter} into the expression above yields:


\begin{equation}
    \alpha^C = \frac{2|C^-|^2 \pi^2(\alpha^{SR}_x+\alpha^{SR}_y)}{4 \pi^2 \Delta^2 + T_0^2(\alpha_x+\alpha^{SR}_y)^2}.
\label{eq:Gamma_proj_eps}
\end{equation}

In this form, $\alpha^C$ explicitly depends on $|C^-|$ and $\Delta$, highlighting that it attains its maximum value near the difference resonance, i.e., when the fractional tune separation vanishes $\Delta \rightarrow 0$.

Under the same assumptions, Eq.~\eqref{eq:eps_ss} reduces to:
\begin{equation}
\begin{split}
    \varepsilon_x = & \frac{\varepsilon_0}{1 + \kappa\frac{\alpha^{SR}_y}{\alpha^{SR}_x}}, \\
    \varepsilon_y = & \frac{\kappa \varepsilon_0}{1 + \kappa\frac{\alpha^{SR}_y}{\alpha^{SR}_x}}.
\label{eq:eps_ss_coupling}
\end{split}
\end{equation}

For a storage ring without combined-function magnets, such as the third-generation light source BESSY II characterized by the damping partition numbers $(1,\,1,\,2)$, full coupling ($\kappa = 1$) would lead to $\varepsilon_x = \varepsilon_y = \varepsilon_0/2$ as $\alpha_x^{SR} = \alpha_y^{SR}$.
In that special case, one recovers the well-known expressions~\cite{guignard_betatron_1995}:
\begin{equation}
\begin{split}
    \varepsilon_x = & \frac{\varepsilon_0}{1 + \kappa}, \\
    \varepsilon_y = & \frac{\kappa \varepsilon_0}{1 + \kappa}.
\end{split}
\end{equation}

In contrast, BESSY III, whose damping partition numbers are $(2.3,\,1,\,0.7)$, yields $\varepsilon_x = \varepsilon_y \approx 0.7\,\varepsilon_0$ under full coupling.

\section{Benchmark with simulations}

\begin{figure}[!htbp]
\centering
   \includegraphics[width=\linewidth]{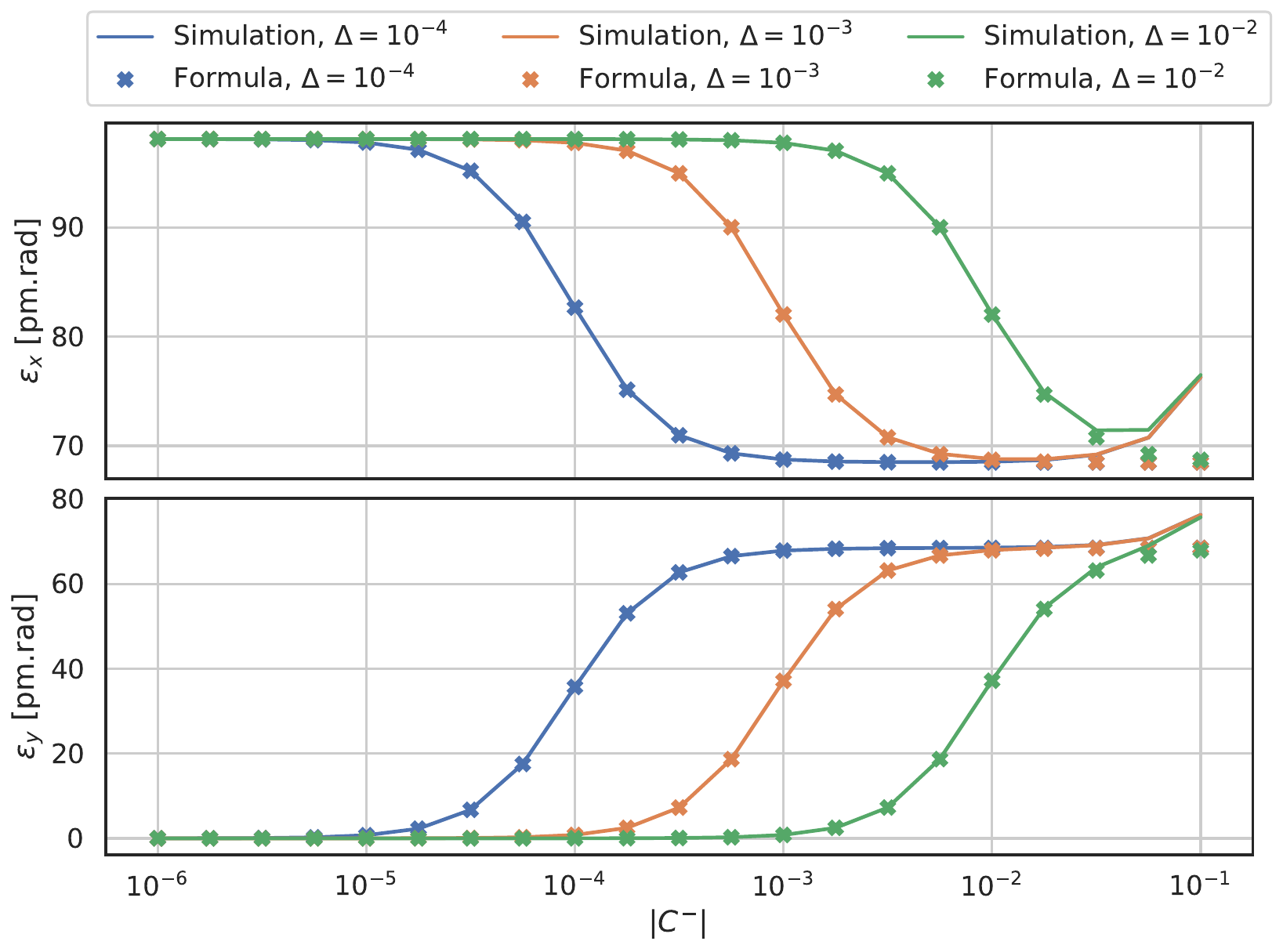}
\caption{Simulated and calculated (using Eq.~\eqref{eq:eps_ss}) projected emittances against the coupling coefficient for different tune separations.}
\label{fig:benchmark_Peter}
\end{figure}

To assess the validity of Eq.~\eqref{eq:proj_eps_peter}, we employ the BESSY~III lattice, whose parameters are listed in Appendix~\ref{app:B}. A skew quadrupole is introduced in a non-dispersive region to control the strength of the betatron coupling, and the working point is shifted to $(44.224 - \Delta/2,\; 12.224 + \Delta/2)$. Using pyAT, the projected emittances are computed for a range of fractional tune separations $\Delta$ by adapting the working point and coupling coefficients $|C^-|$ by varying the skew quadrupole strength. The results of this parameter scan are shown in Fig.~\ref{fig:benchmark_Peter}.

The analytical formulas reproduce the simulated projected emittances with high accuracy over a wide range of $|C^-|$ and $\Delta$. It is worth noting that the equilibrium emittances given by Eq.~\eqref{eq:eps_ss} depend on the uncoupled SR damping rates, $|C^-|$, and $\Delta$, while yielding the same results as the simulations based on the coupled damping partition numbers.
Deviations appear only at large $|C^-|$, where the influence of the sum resonance becomes non-negligible and leads to an emittance blow-up in both planes -- an effect not included in the analytical expressions. Furthermore, the results show that full coupling ($\varepsilon_x = \varepsilon_y$) without a significant presence of the sum resonance is achievable only for $\Delta \lesssim 5 \times 10^{-2}$ based on Fig.~\ref{fig:benchmark_Peter}. For larger $\Delta$, the strong skew quadrupole required to reach full coupling inevitably excites the sum resonance, again causing a blow-up of both emittances. In the following, we therefore choose $\Delta = 10^{-3}$, as it allows full coupling and simultaneously ensures that the working point remains within a tune resolution that can be reliably achieved in a storage ring.

\begin{figure}[!htbp]
\centering
   \includegraphics[width=\linewidth]{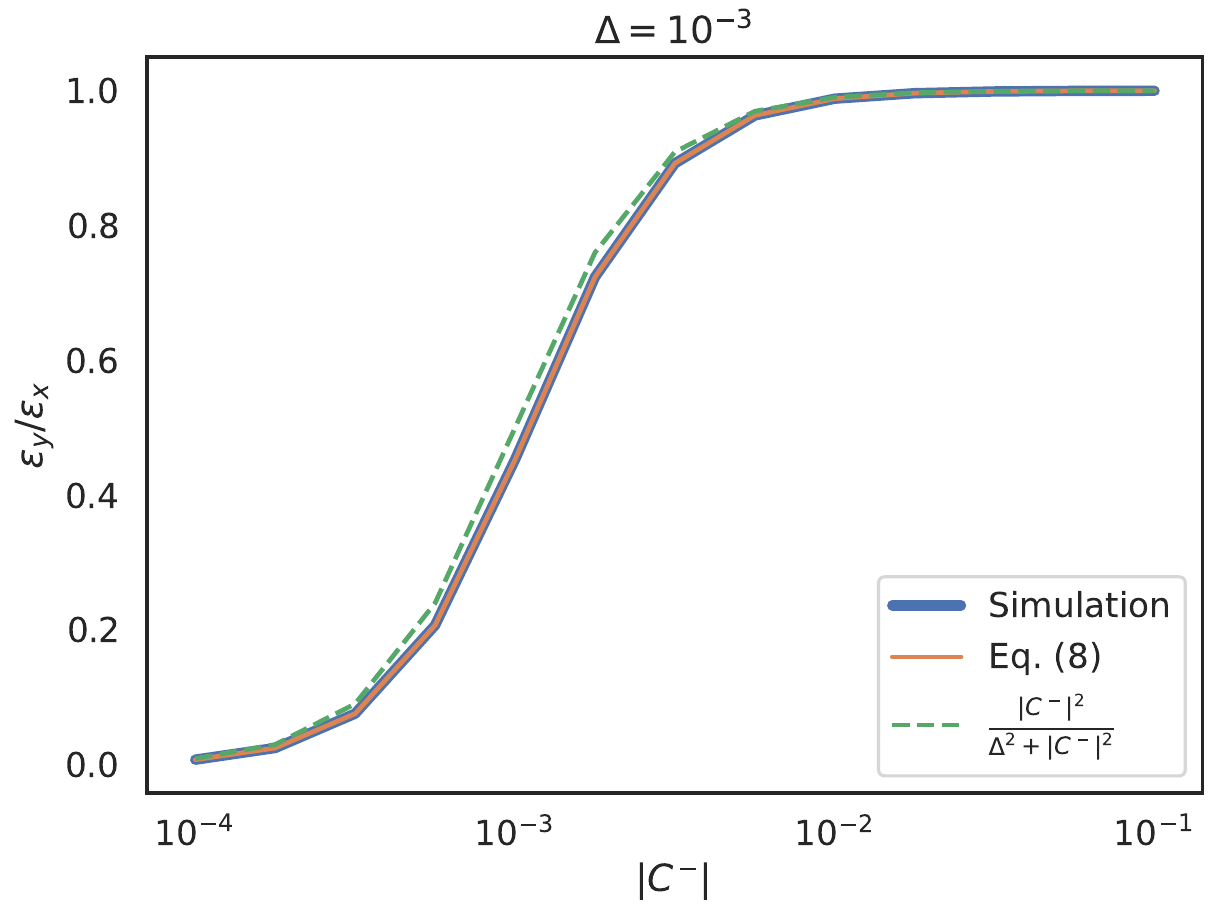}
\caption{Simulated and calculated projected emittance ratios against the coupling coefficient for $\Delta = 10^{-3}$.}
\label{fig:benchmark_ratio}
\end{figure}

A complementary benchmark consists of using the emittance ratio, which can be obtained from Eq.~\eqref{eq:eps_ss}, and comparing it to the classical expression derived in~\cite{guignard_betatron_1995}:
\begin{equation}
    \frac{\varepsilon_y}{\varepsilon_x} = \frac{|C^-|^2}{\Delta^2 +|C^-|^2},
\label{eq:guignard}
\end{equation}

which is valid when $\alpha^{SR}_x = \alpha^{SR}_y$ and neglecting the sum resonance. 
The comparison between both expressions is shown in Fig.~\ref{fig:benchmark_ratio}.

Although Eq.~\eqref{eq:guignard} has been derived for $\alpha^{SR}_x = \alpha^{SR}_y$, it demonstrates an excellent agreement with Eq.~\eqref{eq:eps_ss} over a wide range of $|C^-|$ values, confirming the consistency of the model. The impact of the damping partition numbers is negligible when considering the ratio of the transverse emittances.

Finally, using Eq.~\eqref{eq:guignard}, an alternative simplified expression for $\alpha^C$ can be obtained:

\begin{equation}
    \alpha^C = \frac{2 \alpha^{SR}_y |C^-|^2}{\Delta^2}.
\end{equation}

\section{Generation of vertical emittance}\label{sec:knob}

Vertical emittance in an electron storage ring arises primarily from two main mechanisms: vertical dispersion and betatron coupling. Vertical dispersion is generated by tilts of dipole magnets, transverse misalignments or tilts of quadrupoles, and transverse misalignments of sextupoles located in dispersive regions. In a lattice without misalignments, it can be controlled by inducing vertical orbit offsets in sextupoles using corrector magnets or by powering skew quadrupoles in dispersive regions.
Betatron coupling originates from quadrupole tilts and dedicated skew quadrupoles. When placed in dispersive regions, skew quadrupoles simultaneously introduce betatron coupling and vertical dispersion, whereas non-dispersive skew quadrupoles generate coupling without affecting the vertical dispersion.

In the previous section, betatron coupling was introduced using non-dispersive skew quadrupoles, thereby avoiding the generation of vertical dispersion. In practice however, this approach requires additional magnets in regions where space is limited due to, typically, the presence of insertion devices. An alternative solution consists of using skew-quadrupole coils embedded in existing dispersive sextupoles, while constraining the induced vertical dispersion. The objective is to maximize the contribution of betatron coupling to the vertical emittance while minimizing the dispersion-induced component. This approach allows limiting unwanted vertical dispersion in the insertion devices and effectively decoupling the two effects.

\subsubsection{Betatron coupling knob}

The betatron coupling knob is devised by constraining the vertical dispersion at the bends and insertion devices while maximizing $|C^-|$. The dispersion at the bends is constrained to limit its contribution to the fifth vertical radiation integral, which determines the dispersion-induced vertical emittance:

\begin{equation}
    I_{5,y} = \oint \frac{\mathcal{H}_y}{|\rho|^3} ds,
\end{equation}

where $\mathcal{H}_y = \beta_y D_y'^2 + 2\alpha_y D_y D_y' + \gamma_y D_y^2$, with $\beta_y$, $\alpha_y$, and $\gamma_y$ the vertical Twiss parameters and $\rho$ the bending radius. At the insertion devices, the vertical dispersion is minimized to avoid additional synchrotron radiation from a dispersive vertical orbit.

The relationship between the chosen observables $\overrightarrow{R}$ and the skew-quadrupole strengths $\overrightarrow{K}$ is expressed through the response matrix $\mathbf{M}$, defined as the Jacobian of the system:

\begin{equation}
    \overrightarrow{R} = \mathbf{M} \overrightarrow{K},
\end{equation}

\begin{equation}
\mathbf{M} =
\begin{bmatrix}
\frac{\partial D_{y,1}}{\partial k_1} & \frac{\partial D_{y,1}}{\partial k_2} & \cdots & \frac{\partial D_{y,1}}{\partial k_j} \\
\frac{\partial D_{y,2}}{\partial k_1} & \frac{\partial D_{y,2}}{\partial k_2} & \cdots & \frac{\partial D_{y,2}}{\partial k_j} \\
\vdots & \vdots & \ddots & \vdots \\
\frac{\partial D_{y,i}}{\partial k_1} & \frac{\partial D_{y,i}}{\partial k_2} & \cdots & \frac{\partial D_{y,i}}{\partial k_j} \\
\frac{\partial |C^-|}{\partial k_1} & \frac{\partial |C^-|}{\partial k_2} & \cdots & \frac{\partial |C^-|}{\partial k_j}
\end{bmatrix},
\end{equation}

where $i$ denotes the index of an observable and $j$ the index of a skew quadrupole corrector. 

Both local and global quantities may be included in the response matrix depending on the constraints and objectives of the knob. For instance, $D_y$ may be replaced by the $\mathcal{H}_y$, or $|C^-|$ may be changed with the RDT $|f_{1001}|$ if finer control of the coupling is required. The approach shares similarities with the one developed in~\cite{breunlin_improving_2016}, where the goal was to control the vertical emittance through vertical dispersion bumps.

To further enhance the contribution to $|C^-|$ while keeping the skew quadrupole strengths small~\cite{lee_concurrent_2023}, the skew gradients can be set according to:
\begin{equation}
    k_s \propto \sin{\left(\phi_x - \phi_y - 2\pi \Delta\frac{s}{L} \right)},
\end{equation}
to match the phase dependence of the integrand in Eq.~\eqref{eq:C-}. Under this condition, the sine term remains positive and systematically contributes to the coupling coefficient. 

This constraint can be applied to the response matrix by multiplying $\mathbf{M}$ by a diagonal matrix $\mathbf{F}$, whose elements are defined as:

\begin{equation}
    F_{ii} = \sin{\left(\phi_{x,i} - \phi_{y,i} - 2\pi \Delta\frac{s_i}{L} \right)}.
\label{eq:phase_weight}
\end{equation}

Consequently, the modified response matrix is thus given by $\widetilde{\mathbf{M}} = \mathbf{M} \cdot \mathbf{F}$. Finally, the matrix $\widetilde{M}$ is normalized so that all rows have comparable magnitudes (due to the concurrent use of $D_y$ and $|C^-|$), and additional weights may be applied to emphasize specific observation points.

The optimal skew quadrupole strengths are then obtained by solving the inverse problem:

\begin{equation}
    \overrightarrow{K} = \widetilde{\mathbf{M}}^{-1} \overrightarrow{R},
\end{equation}

while applying a Tikhonov regularization on the singular values of $\widetilde{M}$ to minimize the required skew quadrupole strengths. An over-regularized response matrix ensures that only the most effective combination of skew quadrupoles is used.

A knob relying on skew quadrupole coils embedded in selected dispersive sextupoles provides effective control of the emittance ratio $\varepsilon_y/\varepsilon_x$ without exciting the sum resonance for a working point of $(44.224 - \Delta/2,\; 12.224 + \Delta/2)$ and a tune separation $\Delta = 10^{-3}$. For a fully coupled beam, the maximum skew quadrupole strength required is approximately $0.1\,\mathrm{m^{-2}}$, which is less than $0.2\%$ of the nominal quadrupole gradient in the arcs. Moreover, the knob introduces only minimal vertical dispersion thanks to the phase weighting (Eq.~\eqref{eq:phase_weight}) method described above. 

In the presence of linear betatron coupling, the transverse particle motion is no longer separable into orthogonal horizontal and vertical oscillations, but is instead described by two normal modes (denoted $u$ and $v$). The normal modes correspond to the eigenvectors of the one-turn transfer matrix and define the natural coordinates in which the transverse motion is decoupled.
These modes are generally tilted with respect to the horizontal and vertical planes, so that the physical motion in each plane arises from a superposition of both modes. As a consequence, additional cross-plane beta functions appear. 
The beta functions in the horizontal and vertical planes are obtained by projecting the mode beta functions, analogously to the projection previously used with the mode emittances.

Using the Mais \& Ripken parametrization~\cite{Borchardt:275408, willeke_methods_1989} and its extension by Lebedev \& Bogacz~\cite{lebedev_betatron_2010}, the beta functions projected back onto the horizontal and vertical planes can be expressed as:


\begin{equation}
\begin{split}
    \beta_x = & \frac{\beta_{1x} \mathcal{E}_{u} + \beta_{2x} \mathcal{E}_{v}}{\varepsilon_x}, \\
    \beta_y = & \frac{\beta_{1y} \mathcal{E}_{u} + \beta_{2y} \mathcal{E}_{v}}{\varepsilon_y},
\end{split}
\end{equation}

where $\beta_{ij}$ denote the principal and non-principal beta functions, and $\mathcal{E}_{u}$ and $\mathcal{E}_{v}$ are the corresponding mode emittances. In the absence of coupling, the modes align with the transverse planes, yielding $\beta_x = \beta_{1x}$ and $\beta_y = \beta_{2y}$.

\begin{figure}[!htbp]
\centering
   \includegraphics[width=\linewidth]{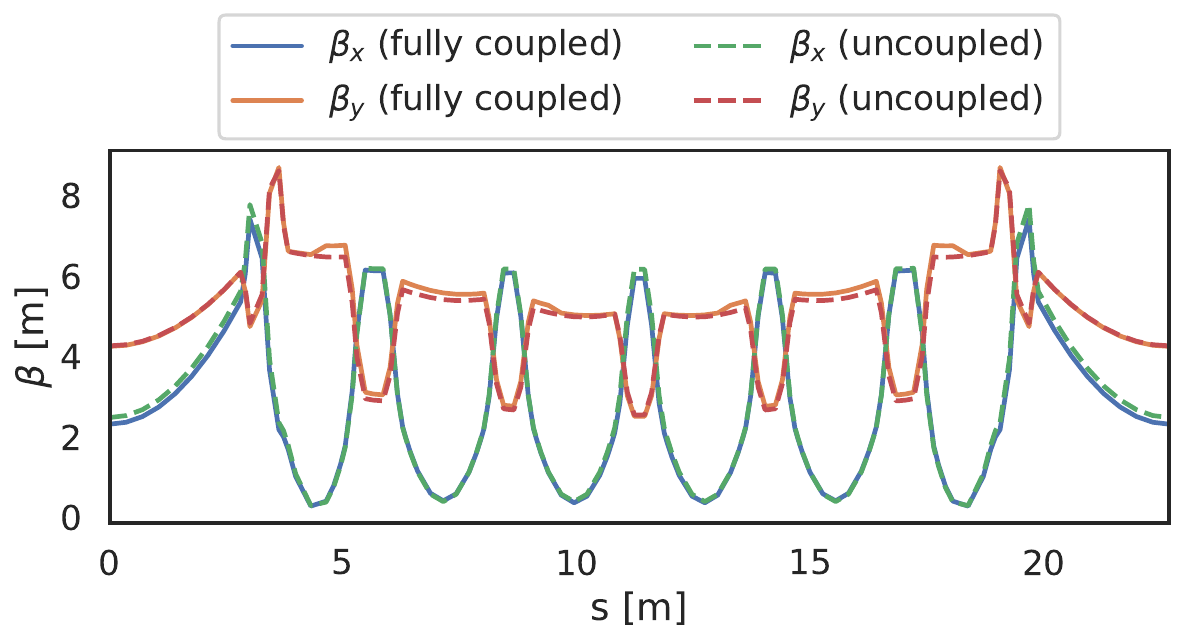}
   \includegraphics[width=\linewidth]{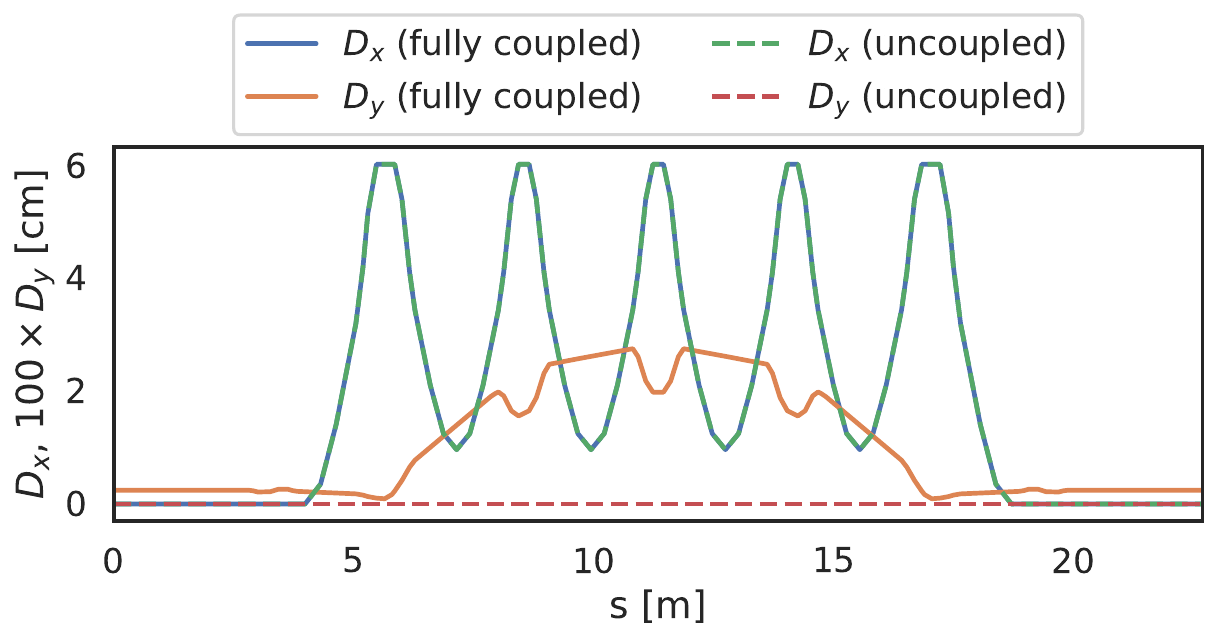}
\caption{Beta functions and transverse dispersion along a superperiod in the uncoupled and fully coupled cases (using the betatron coupling knob), for $(Q_x, Q_y) = (44.224 - \Delta/2,\; 12.224 + \Delta/2)$ and $\Delta = 10^{-3}$.}
\label{fig:beta_functions}
\end{figure}

The resulting projected beta functions and vertical dispersion over a superperiod in the uncoupled and fully coupled cases are shown in Fig.~\ref{fig:beta_functions}. The projected beta functions are almost equivalent to the uncoupled ones. Moreover, the vertical dispersion remains below $25\,\mu\mathrm{m}$ in the straight sections, the regions where the insertion devices will be installed and characterized by a zero horizontal dispersion. The vertical dispersion value is expected to remain below the residual dispersion arising from random sextupole tilts. In comparison, the horizontal dispersion reaches approximately $6\,\mathrm{cm}$ in the arcs. According to this scheme, the vertical emittance is dominated by emittance sharing, with only a minor contribution from vertical dispersion.


\subsubsection{Vertical dispersion knob}

Alternatively, vertical emittance can be generated using a knob that introduces vertical dispersion by powering specific vertical dipolar correctors embedded in the dispersive sextupoles. These correctors are part of the slow orbit feedback scheme~\cite{joly:ipac25-wepm023}, normally used to compensate closed-orbit distortions. When excited with the proper strengths, they generate vertical orbit offsets at the sextupole locations, which in turn produce vertical dispersion through the feed-down effect.

The dispersion-based knob is implemented using the same SVD approach as the betatron coupling knob, but now acting on the vertical dipolar correctors to control $D_y$ at the bends and maximize its contribution to the vertical radiation integral $I_{5,y}$. As in the previous case, the vertical dispersion in the straight sections is constrained to remain minimal, and the overall impact on the optical functions is reduced following the strategy described in~\cite{breunlin_improving_2016}.

\begin{figure}[!htbp]
\centering
   \includegraphics[width=\linewidth]{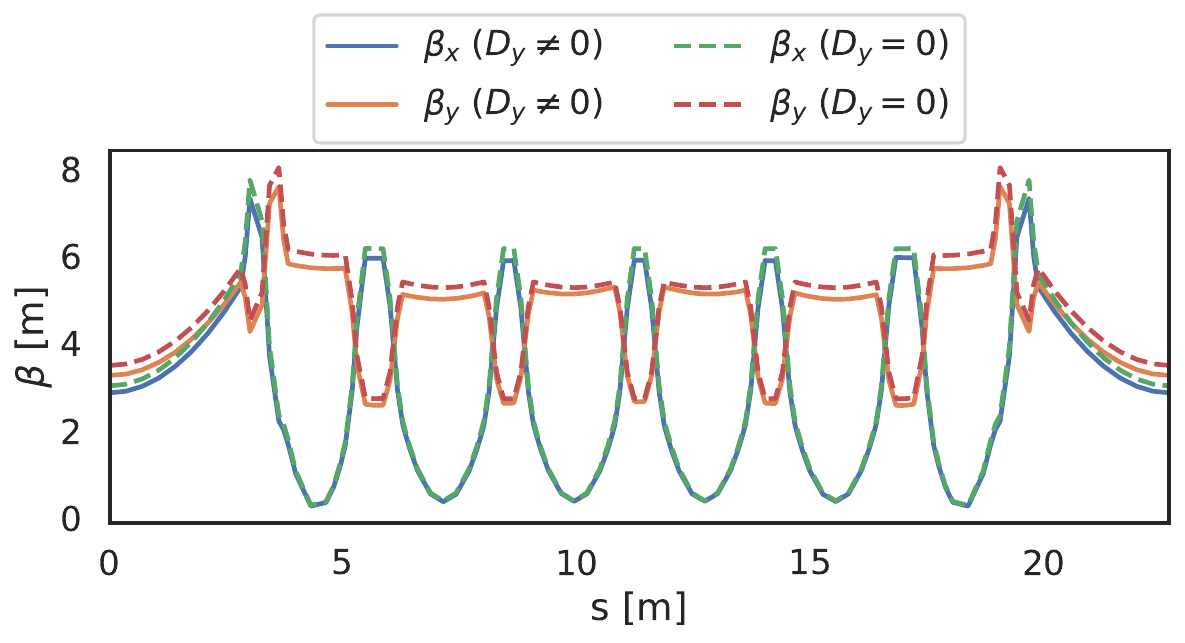}
   \includegraphics[width=\linewidth]{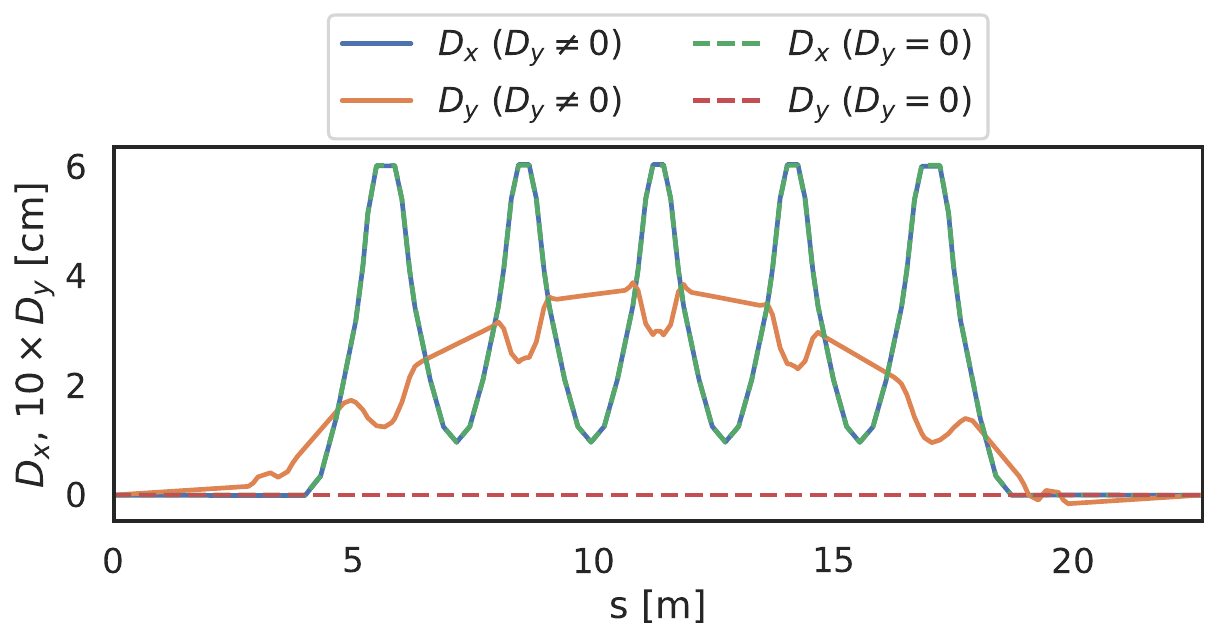}
\caption{Beta functions and transverse dispersion along a superperiod in the uncoupled and lightly coupled cases (using the vertical dispersion knob), for $\varepsilon_y / \varepsilon_x = 0.1$.}
\label{fig:beta_functions2}
\end{figure}

The resulting optical functions for a scenario yielding an emittance ratio $\varepsilon_y / \varepsilon_x = 0.1$ are shown in Fig.~\ref{fig:beta_functions2}. The generated vertical dispersion is approximately a factor of 15 smaller than the horizontal dispersion. Although the knob can be pushed to achieve larger emittance ratios, doing so leads to significant distortions of the beta functions and of the horizontal dispersion. Therefore, it is preferable to use this knob to apply small adjustments to the vertical emittance, in which case only the vertical dispersion differs.

The knobs described above allow the control of vertical emittance through either betatron coupling or vertical dispersion. While their resulting zero-current emittances are defined from the lattice optical functions, IBS will modify the steady-state emittances. Both knobs were used to investigate the interplay between IBS and either the betatron coupling or the vertical dispersion. To do so, we now extend the system of coupled ODEs defined in Eq.~\eqref{eq:heuristic_eps} to include the impact of IBS and, in turn, study the time evolution of the emittances in the presence of these competing mechanisms.

\section{Emittance evolution with IBS and betatron coupling}

The time evolution of the projected emittances in the presence of SR damping, QE, IBS, and betatron coupling is described by the following system of coupled differential equations:
\begin{equation}
\begin{split}
    \frac{d \varepsilon_x}{dt} = & -\alpha^C \left(\varepsilon_x - \varepsilon_y \right) - 2\alpha^{SR}_x \left( \varepsilon_x - \varepsilon_{x,0}\right) + 2\alpha^{IBS}_x \varepsilon_x, \\
    \frac{d \varepsilon_y}{dt} = & -\alpha^C \left(\varepsilon_y - \varepsilon_x \right) - 2\alpha^{SR}_y \left( \varepsilon_y - \varepsilon_{y,0}\right) + 2\alpha^{IBS}_y \varepsilon_y, \\
    \frac{d \varepsilon_z}{dt} = & - 2\alpha^{SR}_z \left( \varepsilon_z - \varepsilon_{z,0}\right) + 2\alpha^{IBS}_z \varepsilon_z, \\
\end{split}
\label{eq:eps_rate_IBS_Gamma}
\end{equation}
where $\alpha^{IBS}_x, \alpha^{IBS}_y, \alpha^{IBS}_z$ are respectively the horizontal, vertical, and longitudinal IBS growth rates.
The steady-state emittances are obtained by setting $d\varepsilon_x/dt = d\varepsilon_y/dt = d\varepsilon_z/dt = 0$, which yields:
\begin{equation}
\begin{split}
    \varepsilon_x & = \frac{2\alpha^{SR}_x\left(\alpha^{C}+2\alpha^{SR}_y-2\alpha^{IBS}_y\right)\varepsilon_{x,0} + 2\alpha^{SR}_y\alpha^{C} \varepsilon_{y,0}}{\left(\alpha^{C}+2\alpha^{SR}_x-2\alpha^{IBS}_x\right)\left(\alpha^{C}+2\alpha^{SR}_y-2\alpha^{IBS}_y\right) - \left(\alpha^{C}\right)^{2}}, \\
    \varepsilon_y & = \frac{2\alpha^{SR}_y\left(\alpha^{C}+2\alpha^{SR}_x-2\alpha^{IBS}_x\right) \varepsilon_{y,0} + 2\alpha^{SR}_x\alpha^{C} \varepsilon_{x,0}}
    {\left(\alpha^{C}+2\alpha^{SR}_x-2\alpha^{IBS}_x\right)\left(\alpha^{C}+2\alpha^{SR}_y-2\alpha^{IBS}_y\right) - \left(\alpha^{C}\right)^2}, \\
    \varepsilon_z & = \frac{\varepsilon_{z,0}}{1-\alpha^{IBS}_z/\alpha^{SR}_z}.
\end{split}
\end{equation}

These equations treat simultaneously the emittance sharing induced by betatron coupling (term $\propto \alpha^C$), the emittance growth due to IBS (term $\propto \alpha^{IBS}_{x,y}$), and the emittance damping due to SR (term $\propto \alpha^{SR}_{x,y}$).

In contrast to the usual approach~\cite{zap}, the emittance sharing is not assumed to occur on a timescale much shorter than IBS. Previously, the steady-state emittances were obtained by setting the initial conditions to the equilibrium emittances with betatron coupling, defined in Eq.~\eqref{eq:eps_ss_coupling}. Then, the coupled system of differential equations (Eq.~\eqref{eq:eps_rate_IBS_Gamma}) was solved neglecting the term proportional to $\alpha^C$, and manually enforcing a chosen emittance ratio at each integration step.

In the proposed formulation, once the desired emittance ratio is obtained by setting the appropriate values of $|C^-|$ and $\Delta$, the system naturally preserves this ratio without the need for additional constraints. 
In addition to the emittance sharing induced by betatron coupling, a finite vertical emittance arising from vertical dispersion is also accounted for through the term $\varepsilon_{y,0}$. 

Once betatron coupling or vertical dispersion is introduced in a lattice, the SR damping rates $\alpha^{SR}_{x,y,z}$ are modified through the redistribution of the transverse damping partition numbers, progressively approaching $(J_x + J_y)/2$ in the fully coupled limit.
In contrast, the IBS growth rates $\alpha^{IBS}_{x,y,z}$ optics' contribution remain mostly unchanged, as the projected beta functions are close to the uncoupled ones. This approximation holds as long as the working point sits near the difference resonance and the coupling remains in the perturbative regime. For larger skew quadrupole strengths, however, the projected beta functions are significantly modified, which in turn impacts the IBS growth rates.
A formula describing the redistribution of the transverse damping partition numbers in the presence of betatron coupling is provided in Appendix~\ref{app:D}. 

\begin{figure}[!htbp]
\centering
   \includegraphics[width=\linewidth]{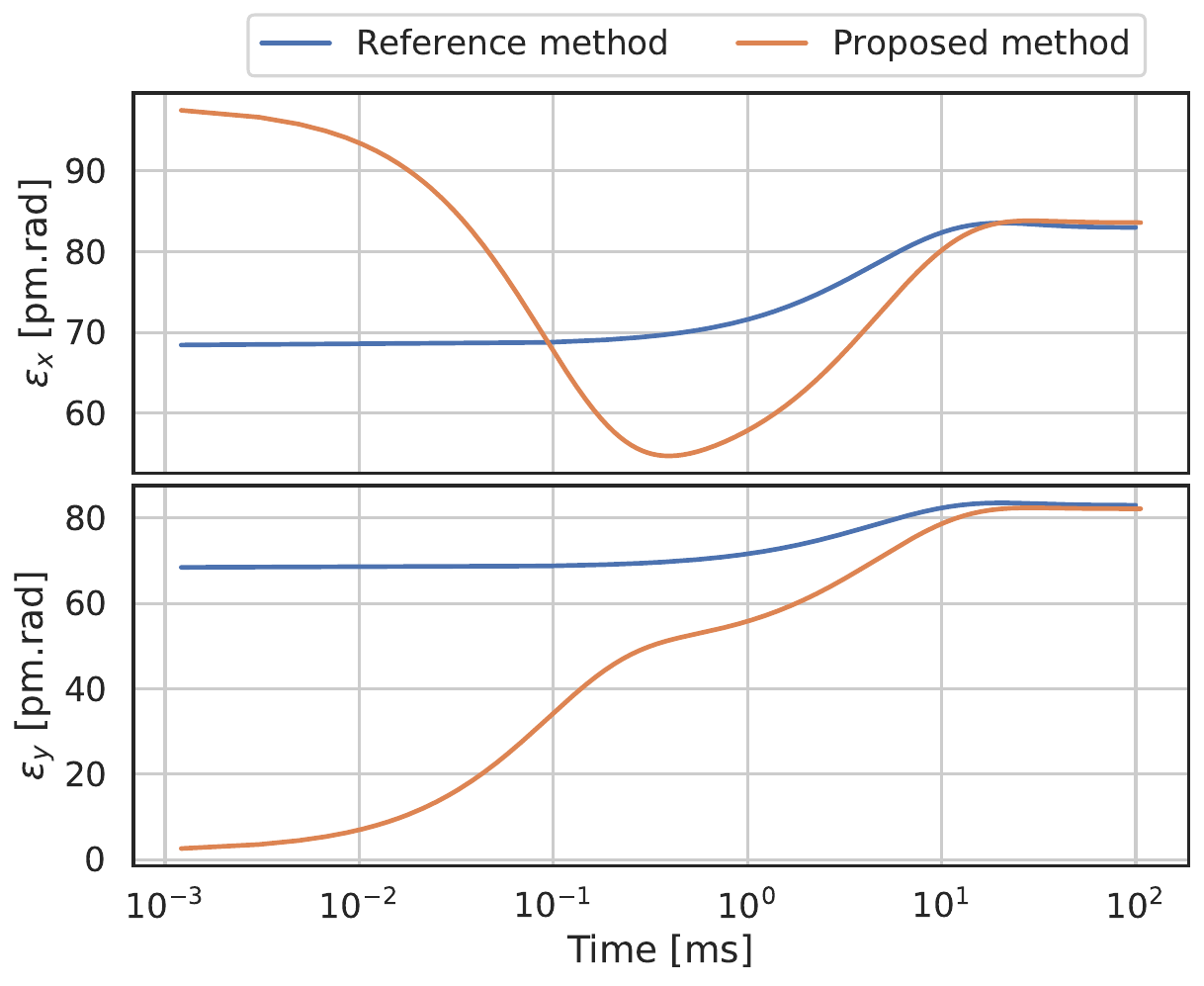}
\caption{Time evolution of the transverse projected emittances obtained with the reference and proposed methods.}
\label{fig:new_impl}
\end{figure}

The proposed approach is benchmarked against the reference implementation~\cite{zap, xsuite, borland_elegant_2000} in the case of full coupling, as shown in Fig.~\ref{fig:new_impl}. While the comparison is performed using projected emittances, Xsuite internally calculates eigenemittances. The results shown were therefore obtained by first converting the eigenemittances into projected emittances using the relations provided in Appendix~\ref{app:C}, and subsequently solving Eq.~\eqref{eq:eps_rate_IBS_Gamma}. The steady-state emittances obtained with both approaches converge to a similar value as they rely on the same $\alpha^{SR}$ and $\alpha^{IBS}$. However, the proposed method provides a more realistic description of the transient dynamics.

\begin{figure}[!htbp]
\centering
   \includegraphics[width=\linewidth]{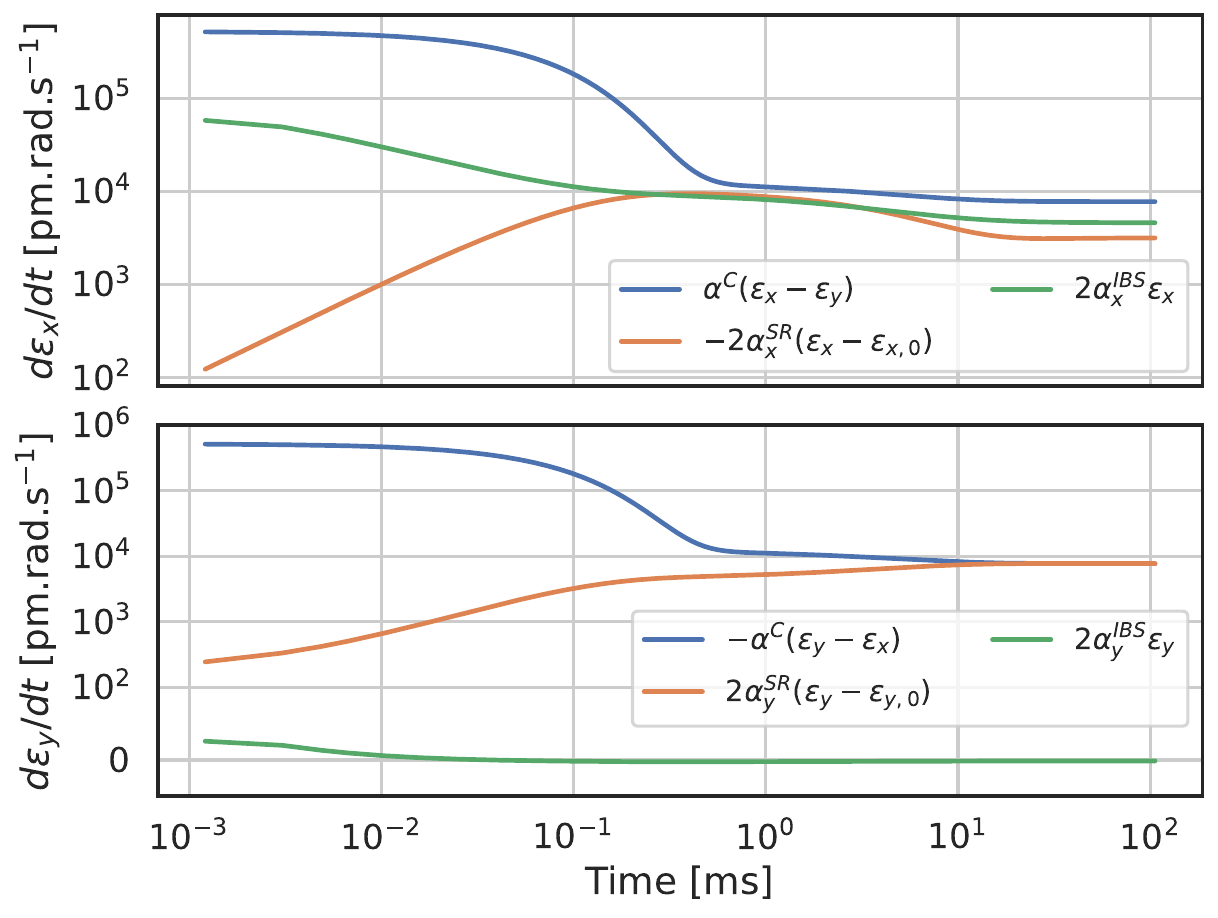}
\caption{Contributions of each term in Eqs.~\eqref{eq:eps_rate_IBS_Gamma} to the horizontal and vertical projected emittance evolution.}
\label{fig:contrib}
\end{figure}

The reference approach requires the initial conditions to be the equilibrium emittances in the presence of SR and betatron coupling, whereas the proposed approach starts from the natural emittances. As shown in Figs.~\ref{fig:new_impl} and~\ref{fig:contrib}, the emittance initially decreases in the horizontal plane and increases in the vertical one due to betatron coupling. Simultaneously, the contribution from IBS drops significantly as the one from SR rises. Once a round beam is reached, all three effects slowly balance each other until the emittances reach their steady-state values. The vertical emittance rate is almost entirely governed by the effects of coupling and SR, with the IBS playing a minor role. As such, the emittance constraint enforced in the reference method is justified as term $2\alpha^{IBS}_y \varepsilon_y$ can be safely neglected.

Using the proposed approach, we can now study the effectiveness of the betatron coupling, vertical dispersion, and vertical excitation at mitigating the effect of IBS.

\section{IBS mitigation strategies (betatron coupling, vertical dispersion, and vertical excitation)}

Apart from relying on betatron coupling and vertical dispersion, vertical emittance can also be introduced by exciting the beam vertically using white noise excitation provided by the transverse feedback system~\cite{Gamelin2024-qk}. This method has the advantage of working with an arbitrary working point and without introducing any vertical dispersion. It also leaves the beta functions, dispersions, SR damping rates, and IBS growth rates untouched.
The vertical excitation is modeled by a constraint on the vertical emittance. At each time step of the numerical resolution of Eq.~\ref{eq:eps_rate_IBS_Gamma}, the vertical emittance is reset to the current value of $\kappa \varepsilon_x$. It effectively neglects vertical emittance growth, with the vertical emittance being solely determined by the horizontal one.
Table~\ref{tab:results} summarizes the steady-state horizontal emittances for various emittance ratios, considering an introduction of vertical emittance through vertical excitation, betatron coupling, or vertical dispersion. Generating vertical emittance through vertical dispersion is inherently limited, as it relies on the vertical dipolar correctors and perturbs the optical functions. For this reason, achieving larger emittance ratios with vertical dispersion would require a complete revision of the lattice. Otherwise, significant degradation of the dynamic aperture and momentum acceptance would occur.

\begin{table}[!htbp]
\centering
\begin{ruledtabular}
\begin{tabular}{lccc}
Method & $\kappa$=10\% & $\kappa$=50\% & $\kappa$=100\% \\
\hline
\multicolumn{4}{l}{\textbf{Excitation}} \\
$\varepsilon_x$ [pm\,rad] & 143 & 121 & 114 \\
$\varepsilon_y$ [pm\,rad] &  14 &  60 & 114 \\
\\
\multicolumn{4}{l}{\textbf{Coupling}} \\
$\varepsilon_x$ [pm\,rad] & 139 & 102 &  84 \\
$\varepsilon_y$ [pm\,rad] &  14 &  51 &  84 \\
\\
\multicolumn{4}{l}{\textbf{Dispersion}} \\
$\varepsilon_x$ [pm\,rad] & 147 & --- & --- \\
$\varepsilon_y$ [pm\,rad] &  11 & --- & --- \\
\end{tabular}
\end{ruledtabular}
\caption{Steady-state horizontal and vertical emittances for different methods of generating vertical emittance and various emittance ratios $\kappa$.}
\label{tab:results}
\end{table}

For small values of $\kappa$, all three methods yield comparable steady-state horizontal emittances. Although vertical dispersion contributes to an increase in $\alpha_y^{IBS}$, the resulting vertical steady-state emittance remains lower than with the excitation and coupling methods. Consequently, a slightly larger horizontal emittance blow-up is observed when vertical dispersion is used.

For larger $\kappa$, clearer differences emerge. The emittance-sharing mechanism induced by betatron coupling allows the horizontal emittance to reach smaller values than with vertical excitation. The largest difference is observed in the round beam case, where the horizontal emittance obtained with betatron coupling is approximately 25\% lower than with excitation. Since the vertical emittance is constrained by the horizontal one, the steady-state vertical emittance achieved with vertical excitation is correspondingly larger than with the coupling method.

\section{Discussion}

The presence of betatron coupling gives rise to cross-term beta functions (i.e., $\beta_{2x}, \beta_{1y}$), whose amplitudes increase with coupling strength. As long as the coupling remains a perturbation, these terms are small and can be safely neglected. These terms were accounted for in a previous section by projecting the principal and non-principal beta functions onto the horizontal and vertical axes. Since most IBS growth rate formalisms assume uncoupled optics and neglect such terms, the IBS growth rates predicted in the fully coupled regime may deviate from the actual values. Using a generalized treatment of IBS in the presence of betatron coupling, similar to that proposed in~\cite{nasht_new_2003}, would allow for a more accurate description.

Based on the obtained results, the trade-off between the reduction of the steady-state horizontal emittance and the operational constraints imposed by betatron coupling must be carefully evaluated. The use of skew quadrupoles and the requirement to operate close to the difference resonance restricts the available working-point space and prevents shifting the tunes to resonance-free regions of the tune diagram. This may lead to a reduction of the dynamic aperture and momentum acceptance, although promising methods are currently under investigation~\cite{carla_methods_2025} to overcome these limitations. The target emittance ratio for BESSY III has not yet been definitively established. However, a baseline vertical emittance of approximately~\SI{10}{\pm}~pm\,rad ($\kappa = 10\%$) is currently under consideration, for which vertical excitation appears to be the most suitable approach.

A realistic prediction of the steady-state emittances in a light source further requires accounting for longitudinal bunch lengthening due to HHC and potential-well distortion, and possibly additional collective effects such as space charge. The ODEs approach reaches its limitations in this context, as it cannot model these phenomena. Therefore, a consistent description requires macroparticle tracking simulations to include all relevant effects.

\section{Conclusion}

In this work, we have extended the ODE-based framework to describe both the steady-state and time evolution of beam emittances in fourth-generation light sources in the simultaneous presence of synchrotron radiation SR, QE, betatron coupling, vertical dispersion, and IBS. In contrast to the conventional approach, the proposed formulation consistently incorporates arbitrary damping partition numbers, their redistribution induced by betatron coupling, and the resulting modifications of the SR damping rates.
We then detailed two practical methods to generate vertical emittance. A carefully optimized configuration of sextupoles in the arcs enables the introduction of betatron coupling while maintaining negligible vertical dispersion. Alternatively, vertical dipole correctors originally dedicated to slow orbit feedback can be repurposed to induce vertical dispersion without coupling, through controlled vertical orbit offsets in the sextupoles.
Both strategies were implemented for the BESSY III lattice in order to evaluate their impact on the steady-state emitances. Despite the redistribution of damping partition numbers induced by betatron coupling, the projected beta functions remain close to their uncoupled counterparts, resulting in nearly invariant IBS growth rates with respect to the coupling strength. Consequently, the steady-state emittances obtained with the new method are consistent with those from the reference approach, while accurately describing the transient behaviors. A comparison of the different mechanisms (i.e., betatron coupling, vertical excitation, and vertical dispersion) highlights the distinct trade-offs associated with each method.

\section{ACKNOWLEDGEMENTS}
The author would like to thank Peter Kuske for the enlightening discussion on betatron coupling and for providing Eq.~\eqref{eq:proj_eps_peter}. This work was supported by the Helmholtz Association and the Federal Ministry of Research, Technology and Space (BMFTR).

\appendix

\section{Emittance exchange (sum resonance)}\label{app:A}

Eqs.~\eqref{eq:heuristic_eps} can be extended to account for both the difference and sum resonances by introducing the difference and sum coupling rates $\alpha^{C_-}$, $\alpha^{C_+}$:

\begin{equation}
\begin{split}
    \frac{d \varepsilon_x}{dt} = & -\alpha^{C_-} \left(\varepsilon_x - \varepsilon_y \right) + \alpha^{C_+} \left(\varepsilon_x + \varepsilon_y \right) - 2\alpha^{SR}_x \left( \varepsilon_x - \varepsilon_{x,0}\right), \\
    \frac{d \varepsilon_y}{dt} = & -\alpha^{C_-} \left(\varepsilon_y - \varepsilon_x \right) + \alpha^{C_+} \left(\varepsilon_x + \varepsilon_y \right) - 2\alpha^{SR}_y \left( \varepsilon_y - \varepsilon_{y,0}\right).
\end{split}
\end{equation}

The final solution is easily deduced by rewriting the previous solution while setting $\alpha^{C} = \alpha^{C_-} - \alpha^{C_+}$:

\begin{equation}
\begin{split}
    \varepsilon_{x,ss} = & \frac{2\varepsilon_{x,0} \alpha^{SR}_x \left(\alpha^{C_+} - \alpha^{C_-} - 2\alpha^{SR}_y\right) - 2\varepsilon_{y,0} \alpha^{SR}_y \left(\alpha^{C_+} + \alpha^{C_-} \right)}{\left(\alpha^{C_+} - \alpha^{C_-} - 2\alpha^{SR}_x\right) \left(\alpha^{C_+} - \alpha^{C_-} - 2\alpha^{SR}_y\right) - \left(\alpha^{C_+} + \alpha^{C_-} \right)^2}, \\
    \varepsilon_{y,ss} = & \frac{2\varepsilon_{y,0} \alpha^{SR}_y \left(\alpha^{C_+} - \alpha^{C_-} - 2\alpha^{SR}_x\right) - 2\varepsilon_{x,0} \alpha^{SR}_x \left(\alpha^{C_+} + \alpha^{C_-} \right)}{\left(\alpha^{C_+} - \alpha^{C_-} - 2\alpha^{SR}_x\right) \left(\alpha^{C_+} - \alpha^{C_-} - 2\alpha^{SR}_y\right) - \left(\alpha^{C_+} + \alpha^{C_-} \right)^2},
\end{split}
\end{equation}

\section{BESSY III parameters}\label{app:B}

\begin{table*}
\begin{ruledtabular}
    \begin{tabular}{lc}
       \textbf{Parameter}                                       & \textbf{BESSY III}      \\
       \hline
       Energy, $E$                                             & \SI{2.5}{GeV}         \\
       \hline
       Horizontal zero-current emittance, $\varepsilon_{x,0}$  & \SI{98}{pm\,rad}      \\
       \hline
       Circumference, $C$                                & \SI{367.36}{m}        \\
       \hline
       Stored Current (612 bunches), $I$                  & \SI{300}{mA} \\
       \hline
       Stored Charge per bunch, $C_B$,                     & \SI{0.6}{nC} \\
       \hline
       Tunes, $Q_x$, $Q_y$                                  & 43.72, 12.79          \\
       \hline
       Momentum compaction factor, $\alpha_c$                & 1.35$\cdot10^{-4}$  \\
       \hline
       Damping numbers $J_x, J_y, J_z$                     & 2.313, 1, 0.687       \\
       \hline
       Damping times, $\tau_x$, $\tau_y$, $\tau_z$,                                & 9.3, 21.5, 31.3 ms      \\
       \hline
       RMS bunch length, $\sigma_z$,                           & \SI{2.4}{mm} \\
       \hline
       RMS energy spread, $\sigma_\delta$                                       & $9.8 \times 10^{-4}$               \\
       \hline
       RF frequency, $f_{RF}$                                        & \SI{500}{MHz}         \\
       \hline
       RF voltage, $V_{RF}$                                          & \SI{2}{MV} \\
       \hline
       Harmonic number, $h$                                     & 612                      \\
       \hline
       Revolution time, $T_0$                         & \SI{1.225}{\mu s}     \\
   \end{tabular}
   \label{tab:b3para}
   \caption{Storage ring parameters of BESSY III.}
\end{ruledtabular}
\end{table*}

\section{Conversion mode emittances}\label{app:C}

In the presence of betatron coupling, the transverse one-turn map can be diagonalized into two normal modes, whose (mode) emittances are $\mathcal{E}_u$ and $\mathcal{E}_v$. The projected emittances $\varepsilon_x$ and $\varepsilon_y$ follow from the projection of the eigenvectors of the one–turn map onto the beam reference frame. Using the formalism of~\cite{franchi_vertical_2011}, the projected emittances evaluated at a position $s$ can be written as:

\begin{equation}
\begin{split}
    \varepsilon_x(s) & = \sqrt{\left(\mathcal{C}(s)^2 \mathcal{E}_u + U(s)\mathcal{E}_v \right)^2 - \left( V(s) \mathcal{E}_v \right)^2}, \\
    \varepsilon_y(s) & = \sqrt{\left(\mathcal{C}(s)^2 \mathcal{E}_v + U(s)\mathcal{E}_u \right)^2 - \left( V(s) \mathcal{E}_u \right)^2},
\end{split}
\end{equation}
where $U(s) = \mathcal{S}_-(s)^2 + \mathcal{S}_+(s)^2$, $V(s) = 2\mathcal{S}_-(s)\mathcal{S}_+(s)$ and the coefficients $\mathcal{C}$, $\mathcal{S}_-$, and $\mathcal{S}_+$ describe the strength of the betatron coupling. They are functions of the coupling resonance driving terms (RDTs):

\begin{equation}
\begin{split}
    \mathcal{P}(s) & = \sqrt{-|f_{1001}(s)|^2 + |f_{1010}(s)|^2}, \\
    \mathcal{C}(s) & = \cosh{(2P(s))}, \\
    \mathcal{S}_-(s) & = \frac{\sinh{(2\mathcal{P}(s))}}{\mathcal{P}(s)} |f_{1001}(s)|, \\
    \mathcal{S}_+(s) & = \frac{\sinh{(2\mathcal{P}(s))}}{\mathcal{P}(s)} |f_{1010}(s)|.
\end{split}
\end{equation}

The quantities $|f_{1001}(s)|$ and $|f_{1010}(s)|$ represent the amplitudes of the difference coupling resonance ($Q_x - Q_y = 0$) and sum coupling resonance ($Q_x + Q_y = 0$) at a position s, respectively. In most storage rings, the sum resonance is negligible, i.e., $|f_{1010}(s)| \ll 1$. Under this approximation, one finds:
\begin{equation}
\begin{split}
    \mathcal{P}(s) & = i|f_{1001}(s)|, \\
    \mathcal{C}(s) & = \cos{(2|f_{1001}(s)|)}, \\
    \mathcal{S}_-(s) & = \sin{(2|f_{1001}(s)|)}, \\
    \mathcal{S}_+(s) & = 0,
\end{split}
\end{equation}

and the expressions for the projected emittances simplify to a linear combination of the mode emittances:
\begin{equation}
\begin{split}
    \varepsilon_x(s) & = \mathcal{C}(s)^2 \mathcal{E}_u + \mathcal{S}_-(s)^2 \mathcal{E}_v, \\
    \varepsilon_y(s) & = \mathcal{C}(s)^2 \mathcal{E}_v + \mathcal{S}_-(s)^2 \mathcal{E}_u.
\end{split}
\label{eq:mode_proj_eps}
\end{equation}

Similarly, inverting the expressions for the projected emittances gives the mode emittances:
\begin{equation}
\begin{split}
    \mathcal{E}_u & = \frac{\mathcal{C}(s)^2 \varepsilon_x(s) - \mathcal{S}_-(s)^2 \varepsilon_y(s)}{\mathcal{C}(s)^2 - \mathcal{S}_-(s)^2}, \\
    \mathcal{E}_v & = \frac{\mathcal{C}(s)^2 \varepsilon_y(s) - \mathcal{S}_-(s)^2 \varepsilon_x(s)}{\mathcal{C}(s)^2 - \mathcal{S}_-(s)^2}.
\end{split}
\label{eq:proj_mode_eps}
\end{equation}

As $\mathcal{C}(s)$ and $\mathcal{S}_-(s)$ are periodic, s-dependent functions, the projected emittances vary along the ring even though the mode emittances remain invariant.
Nonetheless, the mode emittances can be calculated from the projected emittances evaluated at a given location. Choosing $s=0$ as the reference point, one must determine the coefficients $\mathcal{C}(0)$ and $\mathcal{S}_-(0)$, which requires evaluating $|f_{1001}(0)|$ both inside and outside the resonance stopband.

Following~\cite{franchi_emittance_2007}, the RDT amplitude inside the stopband can be expressed as a function of the coupling coefficient and the fractional tune separation:
\begin{equation}
    \sin^2{\left(4|f_{1001}(0)|\right)} = \frac{|C_0^-|^2}{\Delta^2 + |C^-|^2},
\end{equation}

which leads to:
\begin{equation}
    |f_{1001}(0)| = \frac{1}{4} \arcsin{\left(\frac{|C_0^-|}{\sqrt{\Delta^2 + |C^-|^2}}\right)}.
\end{equation}

\begin{figure}[!htbp]
\centering
   \includegraphics[width=\linewidth]{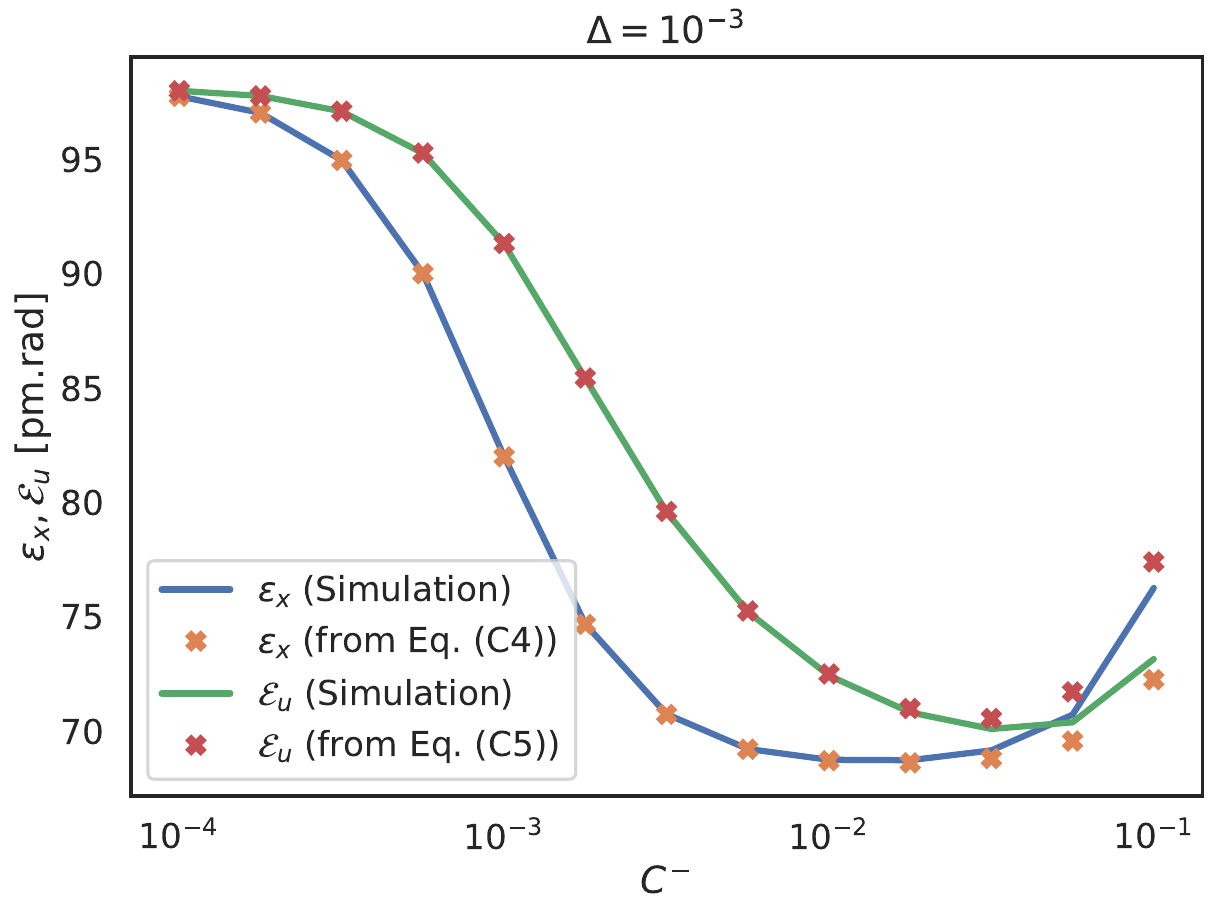}
\caption{Validation of the mode-projected emittance conversion formulas (Eqs.~\eqref{eq:mode_proj_eps} and \eqref{eq:proj_mode_eps}) against the coupling coefficient for $\Delta = 10^{-3}$.}
\label{fig:proof_formula}
\end{figure}

Based on the results of Fig.~\ref{fig:benchmark_Peter} for $\Delta = 10^{-3}$, together with Eqs.~\eqref{eq:mode_proj_eps} and \eqref{eq:proj_mode_eps}, the validity of the formulas can be assessed by transforming the simulated mode emittance into projected emittance and vice versa, as shown in Fig.~\ref{fig:proof_formula}.

The results confirm that the formulas provide an accurate conversion between mode and projected emittances across a wide range of $|C^-|$, except for large values, where contributions from the sum resonance become non-negligible.

\section{Damping partition numbers in the presence of coupling}\label{app:D}

In an electron storage ring, radiation damping is distributed among the transverse and longitudinal planes according to the damping partition numbers $J_x$, $J_y$, and $J_z$. In a coupled lattice, a redistribution of the radiation damping occurs between the transverse planes, leading to different damping partition numbers.

The uncoupled damping partition numbers are defined as:
\begin{equation}
\begin{split}
    J_x = & \oint ds \left(b_{RF} - D_{x} b_{\delta x}\right), \\
    J_y = & \oint ds \left(b_{RF} - D_{y} b_{\delta y}\right),
\end{split}
\end{equation}
where $D_{x}$ and $D_{y}$ are the horizontal and vertical dispersions respectively. $b_{RF}$ is the transverse damping coefficient generated by RF cavities, and $b_{\delta x}, b_{\delta y}$ are the horizontal and vertical damping coefficients due to the bending magnets, respectively, as defined in~\cite{franchi_vertical_2011}.

The damping partition numbers in the presence of coupling can be expressed as a function of the $\mathcal{C}$, $\mathcal{S}_-$, and $\mathcal{S}_+$ coefficients from Appendix~\ref{app:C}:
\begin{equation}
\begin{split}
    J_u = & \oint ds \left[b_{RF} - C^2(0) D_x b_{\delta x} - \left(S_-(0)^2 - S_+(0)^2 \right) D_y b_{\delta y} \right], \\
    J_v = & \oint ds \left[b_{RF} - \left(S_-(0)^2 - S_+(0)^2 \right) D_x b_{\delta x} - C(0)^2 D_y b_{\delta y} \right].
\end{split}
\end{equation}

Neglecting the sum resonance ($\mathcal{S}_+(0) \approx 0$) and rewriting the expressions leads to:
\begin{equation}
\begin{split}
    J_u = & C(0)^2 J_{x} + \left(1 - C(0)^2 \right) J_{y}, \\
    J_v = & C(0)^2 J_{y} + \left(1 - C(0)^2 \right) J_{x}.
\end{split}
\label{eq:coupled_J}
\end{equation}

Thus, the coupled damping partitions can be expressed as a function of the uncoupled $J_x, J_y$ and the $\mathcal{C}(0)$ coefficient, which accounts for the coupling. In addition, the relation $J_x + J_y = J_u + J_v$ is preserved as required by Robinson’s theorem.

\begin{figure}[!htbp]
\centering
   \includegraphics[width=\linewidth]{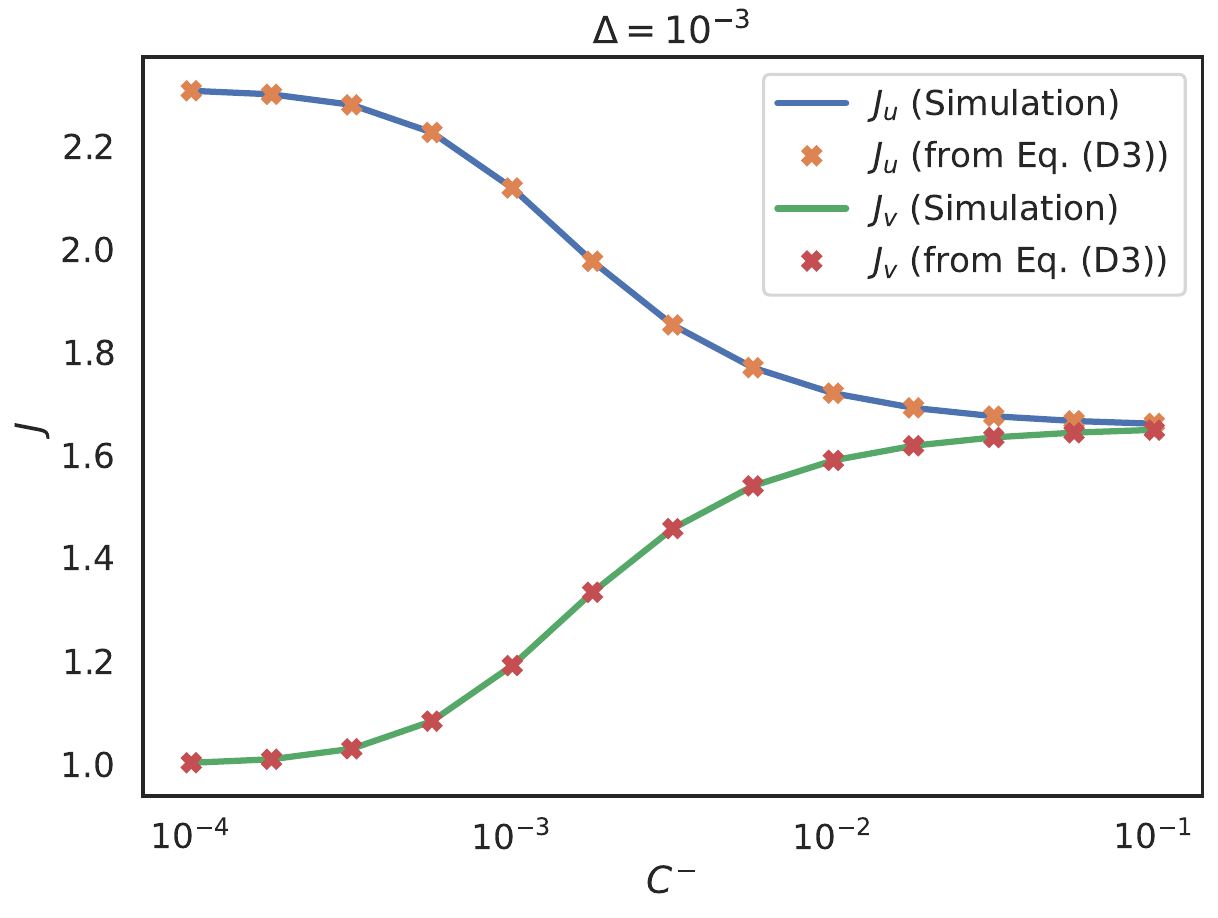}
\caption{Validation of the coupled damping partition number formula (Eq.~\eqref{eq:coupled_J}) against the coupling coefficient for $\Delta = 10^{-3}$.}
\label{fig:proof_J}
\end{figure}

Extracting the coupled damping partition numbers from the simulations corresponding to $\Delta = 10^{-3}$ in Fig.~\ref{fig:benchmark_Peter}, the validity of Eq.~\eqref{eq:coupled_J} can be verified. The result of this comparison can be found in Fig.~\ref{fig:proof_J}.

The coupled damping partition numbers are accurately reproduced by Eq.~\eqref{eq:coupled_J} and show that they become equal for a fully coupled lattice.

\nocite{*}

\bibliography{apssamp}

\end{document}